\documentclass{article}
\usepackage{graphicx}      
\usepackage[authoryear]{natbib}
\setcitestyle{authoryear}
\usepackage{amsmath} 
\usepackage{amssymb}  
\usepackage{url}
\usepackage{authblk}

\usepackage{algorithm}

\usepackage{enumitem}

\usepackage{subcaption}
\newcommand{\Texp}{T_{\mathrm{exp}}}
\newcommand{\Tstop}{T_{\mathrm{stop}}}
\newcommand{\TMPC}{T_{\mathrm{s}}^{\mathrm{MPC}}}
\newcommand{\TCALCMPC}{T_{\mathrm{calc}}^{\mathrm{MPC}}}
\newcommand{\Nu}{N_u}
\newcommand{\Nc}{\Nu}
\newcommand{\Np}{N_p}
\newcommand{\Ts}{T_{\mathrm{s}}}
\newcommand{\Tsmin}{T_{\mathrm{s,min}}}
\newcommand{\JC}{{J^\mathrm{cl}}}
\newcommand{\JCt}{{\tilde{J}^\mathrm{cl}}}

\newcommand{\Nin}{n_{\mathrm{in}}}
\newcommand{\ntheta}{n_\theta}

\newcommand{\nin}{n_u}
\newcommand{\nstate}{n_x}
\newcommand{\nout}{n_y}

{\normalfont}{\normalfont}

\usepackage{color}

\title{Efficient Calibration of Embedded MPC} 

\author[1]{Marco Forgione}
\author[1]{Dario Piga}
\author[2]{Alberto Bemporad}
\affil[1]{IDSIA Dalle Molle Institute for Artificial Intelligence SUPSI-USI, Manno, Switzerland}  
\affil[2]{IMT School for Advanced Studies Lucca,  Lucca, Italy}

\begin{document}
\maketitle

\begin{abstract}                
Model Predictive Control (MPC) is a powerful and flexible design tool of  high-performance controllers for physical systems in 
the presence of input and output constraints. 
A challenge for the practitioner applying MPC is the need of tuning a large number of parameters   such as prediction and control horizons, weight matrices of the MPC cost function, and observer gains, according to different trade-offs. The MPC design task is even more involved when the control law has to be deployed to an embedded hardware unit endowed with limited computational resources. In this case, real-time system requirements limit the complexity of the applicable MPC configuration, engendering additional design tradeoffs and requiring to tune further parameters, such as the sampling time and the tolerances used in the on-line numerical solver. 
To take into account closed-loop performance and real-time requirements, in this paper we tackle the embedded MPC design problem using a global, data-driven, optimization approach
We showcase the potential of this approach by tuning an MPC controller  on two hardware platforms characterized by largely different computational capabilities.
\end{abstract}

\vskip 1em
\noindent\rule{\textwidth}{1pt}
 To cite this work, please use the following bibtex entry:
\begin{verbatim}
@inproceedings{forgione2020efficient,
  title={{E}fficient {C}alibration of {E}mbedded {MPC}},
  author={Forgione, Marco and Piga, Dario and Bemporad, Alberto},
  booktitle={Proc. of the 21st IFAC World Congress, Berlin, Germany},
  year={2020}
}
\end{verbatim}
\vskip 1em
Using the plain bibtex style, the bibliographic entry should look like:\\ \\
M. Forgione, D. Piga, and A. Bemporad. Efficient Calibration of Embedded {MPC}. In \textit{Proc. of the 21st IFAC World Congress, Berlin, Germany}, 2020.

\noindent\rule{\textwidth}{1pt}

\section{Introduction}
Model Prediction Control (MPC) is an advanced control technology that is getting widely popular in different application
domains \citep{borrelli2017predictive}.
The main technical reason of its success is the ability to optimally coordinate inputs and outputs of  multivariable systems in the presence of input/output constraints. 
Besides, the intuitive interpretation of MPC as an optimal controller with respect to a given objective function makes it accessible even to practitioners with limited control  background. 

Nonetheless, calibrating a high-performance MPC controller taking advantage of all the available tuning knobs may still require substantial effort. The challenges normally encountered are: (i) to choose parameters such as prediction and horizon, cost function weight matrices, and observer gains in order to meet desired closed-loop requirements; (ii) to implement the MPC control law on the target hardware platform, ensuring that all computations are performed in real time.

Regarding challenge ($i$),  the final MPC control law is determined by the prediction model, the specified  cost function, and input/output constraints. 
Leaving  aside the constraints, which may be considered direct problem specifications, the practitioner has yet to define the plant model for prediction and the cost function. The model is typically obtained from first-principle laws or estimated from measured data. However, when deriving such a model, a tradeoff emerges between accuracy and complexity, and, most of the times, it is difficult  to decide \emph{a priori} how accurate the model should be in order to achieve  satisfactory closed-loop performance~\citep{FoPiToSa2016,PiFoBe2017}. 
%

As for the MPC cost function, it should represent the underlying engineering or economic objective and --- in some cases --- it could also be a direct specification. However, the MPC cost function is often constrained to have a specific structure, typically a quadratic form of predicted input/output values, to allow the use of very efficient  quadratic programming (QP) numerical optimization
algorithms. Conversely, the true underlying objective resulting from a combination of time- or frequency-domain specifications (or economical considerations) may be formulated more naturally in a different form than a purely quadratic objective. In these cases, the MPC cost function has to be considered as a tuning knob available to achieve the actual closed-loop goal, rather than an exact quantification of the goal itself.
 
Challenge ($ii$) is particularly relevant for fast systems controlled by embedded hardware platforms, such as mobile robots, automotive and aerospace systems, \emph{etc}. Indeed, in embedded systems the on-board computational power is usually limited, \emph{e.g.}, by cost, weight, power consumption, and battery life constraints. Therefore, real-time requirements must be taken into account, further complicating the overall MPC design task.  For instance, due to throughput and memory limitations, the MPC sampling time  cannot be chosen arbitrarily small and the control horizon cannot be chosen arbitrarily large.
Moreover, the low-level design choices of the MPC implementation become crucial and should be taken into account in the design phase. For instance, the engineer should carefully decide whether to go for an explicit MPC approach where the solution is pre-computed offline for all states in a given range \citep{bemporad2002explicit}, or solve the MPC problem on-line by numerical optimization. The first approach is potentially very fast, but requires storing a large lookup table whose size increases with the MPC problem complexity. The applicability of this method is thus limited by the available system memory. Conversely, the second approach has generally a smaller memory footprint, but requires solving a numerical optimization problem on-line. Thus, in the latter case, the MPC problem complexity and the hardware's computational power limit the maximum controller update frequency. 


When the MPC law is computed by numerical optimization (a QP solver in case of linear MPC), the hyperparameters of the optimizer are also MPC tuning knobs, in that they affect solution accuracy and required computation time. 
The overall design of a high-performance MPC must therefore take into account simultaneously high-level aspects related to control systems (model, weights, constraints, prediction horizon, sample time, \emph{etc.})  and low-level aspects of numerical optimization (problem size and solver-related hyperparametes). 

In recent years, data-driven approaches for solving complex engineering tuning problems based on derivative-free global optimization are gaining increasing attention \citep{shahriari2016taking}.
The idea behind these approaches is rather intuitive: the user specifies a search space for the design parameters and the optimization algorithm, based on performance data previously  observed, sequentially suggests the new configurations to be tested, aiming to optimize a user-given performance index. The procedure is iterated until a configuration achieving satisfactory performance is found or the maximum number of available tests has been reached.
This approach has also been popularized as Design and Analysis of Computer Experiments (DACE)~\citep{SWMW89}.
Specialized global optimization algorithms for this task such as Bayesian Optimization
(BO) have been proposed \citep{brochu2010tutorial}. The key feature of these algorithms is their
ability to optimize the objective function with a small number of (possibly noisy) evaluations, without relying on derivative information. 
Recently, optimization-based tuning has been successfully applied to control system design~\citep{roveda2019control, driess2017constrained} and  to choose the MPC prediction model ~\citep{piga2019performance, bansal2017goal}. 

In this paper, we demonstrate the potential of the optimization-based data-driven approach for the joint tuning of high- and low-level MPC parameters in order to optimize a certain closed-loop performance objective, while ensuring that the control law can be  computed in real-time on the hardware platform at hand.   
We apply a  derivative-free global optimization algorithm recently developed by one of the authors~\citep{bemporad2019global}, which has been shown to be very efficient in terms of number of function evaluations  required to solve the global optimization problem.  
We  present the results of our MPC tuning procedure for a simulated cart-pole system on two hardware platforms with very different computational capability, namely a high-end x86-64 workstation and a low-performance ARM-based board (specifically, a Raspberry PI 3, Model B). We show that our tuning procedure can find an MPC configuration that squeezes the maximum performance out of the two architectures for the given control task.

The rest of the paper is organized as follows. The MPC problem formulation and its design parameters are introduced in Section \ref{sec:problem}. Next, the data-driven MPC calibration strategy based on global optimization is described in Section \ref{sec:main} and numerical examples are presented in Section \ref{sec:example}. Finally, conclusions and directions for future research are discussed in Section \ref{sec:conclusions}

\section{Problem formulation}\label{sec:problem}

Let us consider the following nonlinear continuous-time multi-input multi-output dynamical system 
in state-space form
\begin{subequations} \label{eqn:sys}
 \begin{align}
  \dot x &= f(x,u) \\
       y &= g(x,u),
 \end{align}%
 \label{mathcalS}%
\end{subequations}
where $u\in \mathbb{R}^{\nin}$ is the vector of control inputs, $x\in \mathbb{R}^{\nstate}$
the state vector, $y\in \mathbb{R}^{\nout}$ the controlled outputs, $\dot x$ denotes the time derivative of the state $x$, and $f: \mathbb{R}^{\nstate+\nin} \to\mathbb{R}^{\nstate}$ and $g: \mathbb{R}^{\nstate+\nin} \to\mathbb{R}^{\nout}$ are the state and output mappings, respectively. 
Output variables can be collected and used for real-time control at a sampling time $\Ts \geq \Tsmin$, where $\Tsmin$ is the minimum sampling time achievable by the measurement system.

We aim at synthesizing an MPC  (with a state estimator) for~\eqref{eqn:sys}  such that the resulting closed-loop system minimizes a certain closed-loop performance index $\JC$. This performance index is defined as a continuous-time functional $\JC(y_{[0,\;\Texp]}, u_{[0,\;\Texp]})$ in  an experiment of fixed duration  $\Texp$, where $y_{[0,\;\Texp]}$ (resp. $u_{[0,\;\Texp]}$) denotes the output signal $y(t)$ (resp. input signal $u(t)$) within the time interval $[0,\;\Texp]$.  
%
%
As an additional requirement, we must ensure that the control law can be computed in real-time on a given hardware platform. This requirement is translated into the constraint $\TCALCMPC \leq \TMPC$, where $\TMPC$ denotes the MPC sampling time  and $\TCALCMPC$ the worst-case time required to compute the MPC control law on the given platform. 
It is worth remarking that in general the closed-loop performance index $\JC$ is a nonconvex function of the MPC design parameters, that we will defined later. For the sake of generality, $\JC$ has been denoted above as a continuous-time functional over the duration $\Texp$ of the experiment. Thus, $\JC$ does not necessarily correspond to the cost function minimized on-line by MPC. Indeed, the latter is generally defined as a discrete-time quadratic function over a prediction horizon generally shorter than $\Texp$. 

In the following paragraphs, we define the MPC and state estimator design problem, 
along with their tuning parameters. 

\subsection{MPC controller} \label{Sec:outerMPC}
We assume that a continuous-time (possibly parametrized) model $M(\theta^m)$ of~\eqref{mathcalS} is available, where $\theta^m$ represents a vector of adjustable model parameters.
For a given choice of $\TMPC$, $M(\theta^m)$ can be linearized and discretized in time, yielding the discrete-time state-space model 
\begin{subequations}  
\label{eq:SITO}
\begin{align}
x_{t+1}&=A(\TMPC,\theta^m)x_t+B(\TMPC,\theta^m)u_t \label{eq:MPCsysstate}\\
    y_t&=C(\theta^m)x_t+D(\theta^m)u_t \label{eq:MPCsysout}
\end{align}
\end{subequations}
that is used as prediction model for MPC.

At each time $t$ that is an integer multiple of the MPC sampling time $\TMPC$,  MPC solves the  minimization problem
\begin{subequations}  \label{eq:MPC}
	\begin{align}
	&  \nonumber \min_{\left\{u_{t+k|t}\right\}_{k=0}^{\Nu-1}, \epsilon}  \sum_{k=0}^{\Np-1} \left(y_{t+k|t}-y^{\rm{ref}}_{t+k}\right)^\top Q_y\left(y_{t+k|t}-y^{\rm{ref}}_{t+k}\right) +\\ 
	& \nonumber \qquad \qquad+\sum_{k=0}^{\Np-1} \left(u_{t+k|t}-u^{\rm ref}_{t+k}\right)^\top  Q_u \left(u_{t+k|t}-u^{\rm ref}_{t+k}\right) \!+\!\\
	&\qquad \qquad + \sum_{k=0}^{\Np-1} \Delta u_{t+k|t}^\top Q_{\Delta u} \Delta u_{t+k|t}  + Q_\epsilon \epsilon^2 \label{eq:MPCcost}\\
	&  \mathrm{s.t. \ }   \text{model equations } \eqref{eq:MPCsysstate},\eqref{eq:MPCsysout} \\
	& \quad y_{\mathrm{min}} \!-\! V_y\epsilon\leq y_{t+k|t} \leq  y_{\mathrm{max}}+V_y\epsilon, \  k=1,\ldots,\Np  \\
	& \quad   u_{\mathrm{min}}\!-\!V_u\epsilon \leq u_{t+k|t} \leq  u_{\mathrm{max}}+V_u\epsilon, \  k=1,\ldots,\Np \\
	& \quad  \Delta u_{\mathrm{min}} -V_{\Delta u}\epsilon \leq \Delta u_{t+k|t}, \ \    k=1,\ldots,\Np \label{eq:MPC:Ducons1}  \\
	& \quad    \Delta u_{t+k|t} \leq  \Delta u_{\mathrm{max}}+V_{\Delta u}\epsilon, \ \  k=1,\ldots,\Np \label{eq:MPC:Ducons2}\\ 
	& \quad  u_{t+\Nu+j|t}= u_{t+\Nu|t},  \ \  j=1,\ldots, \Np-\Nu, \label{eq:MPCconst}
 	\end{align}
\end{subequations}
where $\Delta u_{t+k|t} =  u_{t+k|t}- u_{t+k-1|t}$;  $\Np$ and $\Nu$ are the prediction and control horizon, respectively; $Q_{y}$, $Q_{u}$, and  $Q_{\Delta u}$ are positive semidefinite weight matrices specifying the MPC cost function; $u_{\rm ref}$ and $y_{\rm ref}$ are the input and output references, respectively;  $Q_{\epsilon}$, $V_y$, $V_u$, $V_{\Delta u}$ are positive constants used to soften the input and output constraints, ensuring that the optimization problem~\eqref{eq:MPC} is always feasible. 
An MPC calibrator would typically adjust $\Np,\Nc, Q_y, Q_u, Q_{\Delta u}$ using a mix of experience and trial-and-error until the desired closed-loop goals are achieved,
fixing the remaining parameters to their default value.
Such a process, in particular in the absence of a deep knowledge of MPC, can be very time consuming and therefore costly.

Several parameterizations may be used to simplify the calibration task. For instance, weight matrices $Q_y$, $Q_u$, and $Q_{\Delta u}$ may be constrained to be diagonal (one of the weights may also be chosen equal to one without loss of generality). 
For notation convenience, we denote by $\theta^c$ the set of all tuning parameters of the MPC problem introduced above.

The solution of the QP problem \eqref{eq:MPC} is computed through numerical optimization. 
We denote by $\theta^s$ the QP solver settings, that we assume can also be adjusted by the calibrator. For instance, important solver parameters are the QP's relative and absolute
feasibility/optimality tolerances for termination. Note that the parameters $\theta^s$ influence both the accuracy of the numerical solution (thus, the performance index $\JC$) and the computation time (thus, $\TCALCMPC$).

\subsection{State estimator}
An estimate of the system state $x_t$ is required to solve the MPC optimization problem~\eqref{eq:MPC}. In this paper, we use a Luenberger observer for state estimation:
\begin{subequations}
\begin{align}
 \hat x_{t+1|t}   &= A \hat x_{t|t-1} + B u_t + L(y_t - C\hat x_{t|t-1})\\
 \hat y_{t+1|t}  &= C \hat x_{t+1|t},
\end{align}
\end{subequations}
where $\hat x_{t|t-1}$ is the state estimate at time $t$ based on observations up to time $t-1$.
We compute the gain $L$ as the standard stationary Kalman filter gain, based on the (linearized) model \eqref{eq:SITO}, assuming positive definite covariance matrices $W_w$ and $W_v$ for the additive process and measurement noise, respectively.
As for the MPC tuning weights, different parametrizations/structures may be used to 
define such covariance matrices. The corresponding parameters are the tuning knobs of the state estimator and are denoted as $\theta^f$.               

\section{Performance-driven parameter tuning}\label{sec:main}

For notation convenience, the design parameters $\theta^m$, $\theta^c$, $\theta^s$, $\theta^f$ introduced in the previous section are collected in the single vector $\theta \in \mathbb{R}^{\ntheta}$.

In this section, we describe how to tune $\theta$ through an experiment-driven approach in order to optimize the closed-loop performance index $\JC(y_{[0,\Texp]},u_{[0,\Texp]}; \theta)$, under the real-time constraint $\TCALCMPC(\theta) \leq \TMPC$.
The overall MPC design task can be formalized as the following constrained global 
optimization problem
\begin{subequations} 
\label{eq:GOP}
	\begin{align} 
	&   \min_{\theta \in \Theta}{\JC(y_{[0,\Texp]},u_{[0,\Texp]}; \theta)} \label{eq:TCALCcst} \\ 
	&  \mathrm{s.t. \ }   \TCALCMPC(\theta) \leq \eta \TMPC. \label{eq:TCALCcnst}
	\end{align}
\end{subequations}
In~\eqref{eq:TCALCcst}, $\Theta \subseteq \mathbb{R}^{\ntheta}$ is the set of admissible values  of the design vector $\theta$. Specifically, in this work, $\Theta$ is a box-shaped region  delimited by lower and upper bounds for each individual parameter.
The constant $\eta$, $0<\eta<1$ in~\eqref{eq:TCALCcnst} takes into account that, in a practical implementation, a fraction of the controller's computation time should be left available for other tasks. 

It is important to stress that neither for $\JC(\theta)$ nor for $\TCALCMPC(\theta)$ a closed-form expression is available.  
Nevertheless, these functions can be evaluated through real experiments or simulations\footnote{In order to evaluate the worst-case computation time $\TCALCMPC$ for a given parameter $\theta$, the control law should be computed on the target hardware directly. In the absence of a platform emulator, this can be done, for instance, with an \emph{hardware-in-the-loop} setup where the target hardware closes the control loop on a simulated system.}.
In the following, we describe the optimization algorithm used to solve problem \eqref{eq:GOP} based on function evaluations of $\JC$ and $\TCALCMPC$. 
One of main strengths of this algorithm is its efficiency in terms of number of required 
function evaluations.



\subsection{Global optimization for parameter selection} \label{sec:GLIS}
First, the constrained optimization problem~\eqref{eq:GOP} is approximated by the following box-constrained problem
\begin{subequations} \label{eqr:probConst}
	\begin{align} 
	&   \min_{\theta \in  \Theta}{\JCt(\theta)},
	\end{align}
where $\JCt(\theta)$ is defined as the original cost $\JC( \theta)$ plus a continuous barrier function $\ell: \mathbb{R} \rightarrow \mathbb{R}^+$ penalizing the violation of the constraint $\TCALCMPC(\theta) \leq \eta \TMPC$, \emph{i.e.}, 
	\begin{align}  \label{eqn:Jtilde}
	&  \JCt(\theta) = \JC( \theta)  + \ell\left(\TCALCMPC(\theta) - \eta \TMPC\right).
	\end{align} 
\end{subequations}

We solve the optimization problem~\eqref{eqr:probConst} by using the approach described
in~\citep{bemporad2019global}, called GLIS (GLobal optimization  based on Inverse distance weighting and radial basis function Surrogates).

The algorithm first runs $\Nin \geq 1$ closed-loop experiments 
for $\Nin$ different values of the vector of controller parameters $\theta_n$, $n=1,\ldots,\Nin$, 
generated randomly using Latin Hypercube Sampling (LHS)~\citep{mckay1979comparison}
within $\Theta$, measuring the corresponding  performance index $\JCt_n$. 
Next, at iteration $n \geq \Nin$ a radial basis function $\hat f_{\rm RBF}$ is fit to the available samples $\{(\theta_1,\JCt_1),\dots, (\theta_n,\JCt_n)\}$. This function $\hat f_{\rm RBF}$ is a surrogate of the (non-quantified) performance index $\JCt$. 
Another function $z:\Theta\to\mathbb{R}$, that promotes the exploration of the set $\Theta$ in areas that
have not been sampled yet and where the empirical variance of $\JCt-\hat f_{\rm RBF}$
is large, is summed to $\hat f_{\rm RBF}$ to define an acquisition function $a:\Theta\to\mathbb{R}$.
The function $a$ (which is very easy to evaluate and even differentiate) is then minimized over $\Theta$ by using a global optimization algorithm,
such as Particle Swarm Optimization (PSO)~\citep{EK95,VV09},
to get a new configuration  $\theta_{n+1}$ 
of MPC parameters to test. A new closed-loop experiment is performed with controller parameters $\theta = \theta_{n+1}$ and the performance index $\JCt_{n+1}$ is measured.
The algorithm iterates  until a stopping criterion is met or a maximum number $n_{\mathrm{max}}$ of iterations is reached.

The main advantage of using the global optimization algorithm GLIS described above  for  solving  the calibration problem  is twofold. First, it is a derivative-free  algorithm, and thus it is particularly convenient since  a closed-form expression of the cost $\JCt$ as a function of the design parameters $\theta$ is not available. Second, it allows us to tune the controller parameters $\theta$ with a smaller number of experiments compared to other existing global optimization methods,
such as PSO, DIRECT (DIvide a hyper-RECTangle)~\citep{Jon09},
Multilevel Coordinate Search (MCS)~\citep{HN99}, Genetic Algorithms (GA)~\citep{Han06},
and usually even than BO as reported in~\citep{bemporad2019global}.



\section{Numerical Example}
\label{sec:example}
\begin{figure}
	\centering
	\includegraphics[width=.4\textwidth]{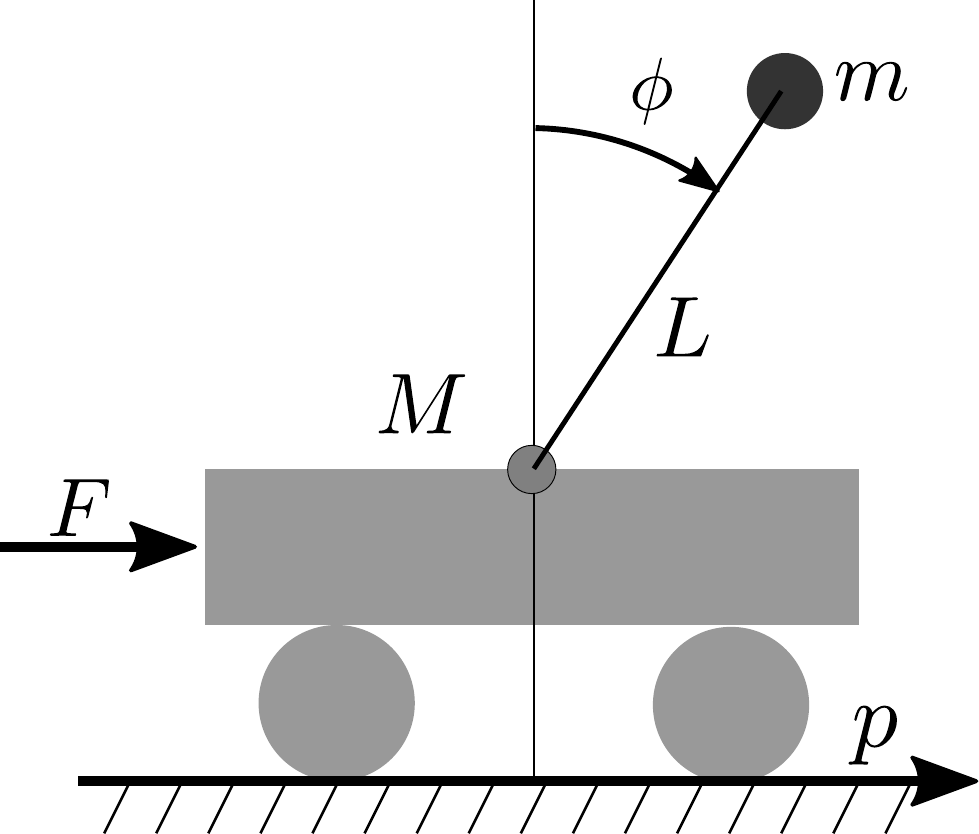} 
	\caption{Schematics of the cart-pole system.}
	\label{fig:pendulum_cart}  
\end{figure}

As a case study, we consider the problem of controlling the cart-pole system depicted in Fig.~\ref{fig:pendulum_cart}.  
We aim at designing an MPC controller that minimizes a given closed-loop performance index $\JC$, while satisfying the constraint~\eqref{eq:TCALCcnst} on the worst-case MPC execution time $\TCALCMPC$ for real-time implementation.
The MPC law is computed using a custom-made Python library that transforms problem~\eqref{eq:MPC} into a standard QP form, which is subsequently solved using the ADMM-based QP solver OSQP \citep{stellato2018osqp}. 
The MPC parameters are tuned via global optimization using the solver GLIS
recalled in Section~\ref{sec:GLIS} and retrieved from \url{http://cse.lab.imtlucca.it/~bemporad/glis}. In particular, package version 1.1  was used with default settings.

The source code generating the results in this paper can be found at \url{https://github.com/forgi86/efficient-calibration-embedded-MPC}. 
A standalone installation version of the MPC library is also available at \url{https://github.com/forgi86/pyMPC} for convenient integration in other projects.

The MPC tuning is performed for two different hardware platforms with remarkably different speed performance:
\begin{itemize}
\item an x86-64 PC equipped with an Intel i5-7300U 2.60~GHz CPU and 32  GB  of  RAM;
\item a Raspberry PI 3 rev. B board equipped with a 1.2~GHz ARM Cortex-A53 CPU and 1~GB of RAM.
\end{itemize}

The Raspberry PI 3 is roughly 10 times slower than the PC in computing the MPC law. This leads to different constraints on the maximum controller complexity and thus on the achievable closed-loop performance. 

\subsection{System description}
The cart-pole  dynamics are governed by the  following continuous-time differential equations which are used to simulate the behavior of the system:  
	\begin{align*} 
	(M+m)\ddot p + mL\ddot\phi \cos\phi - mL \dot \phi ^2 \sin \phi + b\dot p &= F\\
	L \ddot \phi + \ddot p \cos \phi - g \sin\phi + f_\phi\dot \phi &=0 ,
	\end{align*}
where  $\phi\,\mathrm{(rad)}$ is the angle of the pendulum with respect to the upright vertical position, $p\,\mathrm{(m)}$ is the cart position, and $F\,\mathrm{(N)}$  is the input force  on the cart. The following values of the physical parameters are used:  $M=0.5\,\mathrm{kg}$ (cart mass), $m=0.2\,\mathrm{kg}$ (pendulum mass), $L=0.3\,\mathrm{m}$ (rod length), 
$g=9.81\,\mathrm{m/s^{2}}$ (gravitational acceleration), $b=0.1\,\mathrm{N/m/s}$, and $f_\phi=0.1\,\mathrm{m/s}$ (friction terms). 
Measurements of $p$ and $\phi$
are supposed to be collected at a minimum sampling time $\Tsmin=1\,\mathrm{ms}$, and are  corrupted by additive zero-mean white Gaussian noise sources with standard deviation $0.02\,\mathrm{m}$ and $0.01\,\mathrm{rad}$, respectively. The input force $F$ is  perturbed by an additive zero-mean colored Gaussian noise  with standard deviation $0.1\,\mathrm{N}$ and bandwidth $5\,\mathrm{rad/s}$. 
The force $F$ is bounded to the interval  $[-10,\; 10]\, \mathrm{N}$, which is due to actuator saturation, while the cart position $p$ is limited to the interval $[-1.2,\;   1.2]\,\mathrm{m}$, representing the length of the track where the cart moves.

The system is initialized at $[p(0) \;  \dot{p}(0)\; \phi(0)\;\dot{\phi}(0)] = [0 \;  0\; \frac{\pi}{18}\;0]$. In the MPC calibration experiment, the objective is to track a piecewise linear position reference $p^{\mathrm{ref}}$ for the cart passing through the time-position points $\{(0,0)$, $(5,0)$, $(10,0.8)$, $(20,0.8)$, $(25,0)$, $(30,0)$, $(40,0.8)\}$ over an experiment of duration $\Texp=40$ s, while stabilizing the angle $\phi$ to the upright vertical position, \emph{i.e.,} at $\phi=0$. 

The controller is disabled at time $\Tstop < \Texp$ whenever one of the following early termination condition occurs: 
\begin{itemize}
\item pendulum falling ($|\phi| > \frac{\pi}{6})$
\item cart approaching end of the track ($|p| \geq 1.1~\mathrm{m}$)
\item numerical errors in the MPC law computation
\end{itemize}
Similar conditions may be required to ensure safety in the case of real experiments performed on a physical setup. 
In our simulation setting, they are still useful to reduce the computational time as they speed up the test of configurations that are definitely not optimal. 
Furthermore, early termination is explicitly penalized in our closed-loop performance index (see next paragraph), and thus provides useful information for MPC calibration to the global optimization algorithm.

\subsection{Performance index}
The closed-loop performance index $\JCt$ is defined as 
\begin{multline}
\label{eq:JCL_example}
 \JCt =  \ln \left ( \int_{t=0}^{\Texp}10 |p^{\mathrm{ref}}(t) - p(t)| +  30 |\phi(t)|\;dt \right )+ \\
  +  \ell\left(\TCALCMPC - \eta \TMPC\right) + \ell'(\Texp - \Tstop),
\end{multline}  
where the penalty term $\ell$ for real-time implementation of the control law is
\begin{equation}
\label{eq:penalty}
 \ell\!=\! \begin{cases} 
 \!\ln \left(1+ 10^3\frac{\TCALCMPC\!-\!\eta \TMPC}{\eta \TMPC}\right)&\text{if } \TCALCMPC > \eta \TMPC\\
 0 &\text{otherwise},
 \end{cases}
\end{equation}
with $\eta = 0.8$. Another term $\ell'$ is used to penalize early termination conditions, as previously discussed:
\begin{equation}
\label{eq:penalty_time}
 \ell'\!=\! \begin{cases} 
 \!\ln \left(1+ 10^3\frac{\Texp\!-\!\Tstop}{\Texp}\right)&\text{if } \Tstop < \Texp\\
 0 & \text{otherwise}.
 \end{cases}
\end{equation}
The integral in~\eqref{eq:JCL_example} is approximated using samples collected at the fastest sampling time $\Tsmin$. 

\subsection{Control design parameters}
We have different MPC design parameters to tune in order to minimize the performance-driven
objective~\eqref{eq:JCL_example}.

As for the MPC cost function, the positive definite weight matrix $Q_y$ is diagonally parameterized as 
$\bigl( \begin{smallmatrix}
  q_{y_{11}}&0\\ 0&q_{y_{22}}
\end{smallmatrix} \bigr)$, 
where $q_{y_{11}}$, and $q_{y_{22}}$ are design parameters taking real values in the interval $[10^{-16},\   1]$.  Since in the example $n_u=1$, the weights $Q_u$ and $Q_{\Delta u}$  are  scalars.  $Q_{\Delta u}$ is taken as a decision variable and bounded in the range $[10^{-16} ,\  1]$, while $Q_u$ is set to 0. The prediction horizon $\Np$ is an integer  parameter in the range $[5,\ 300]$, while the control horizon $\Nc$ is a fraction $\epsilon_c$ of $\Np$ rounded  to the closest integer,  with design parameter $\epsilon_c$ restricted to the range $[0.3,\ 1]$. Lastly, the MPC sampling time $\TMPC$ is  a design parameter restricted to the range $[1,\ 50]\,\mathrm{ms}$.

In the QP solver, the relative and absolute feasibility/optimality tolerances are tuned. Specifically, the log of two tolerances are parameters in the range $[-7,\ -1]$. 

As for the state estimator, the 4x4 process noise covariance matrix $W_w$ and the 2x2 output noise covariance matrix $W_v$ are diagonally parametrized, similarly to  $Q_y$.

The system dynamics are assumed to be known. Therefore, there is no tunable model parameter in our design problem. 

Finally, MPC is configured with fixed constraints $F_{\max}\!=\!-F_{\min}\!=\!10~\mathrm{N}$ and $p_{\max}\!=\!-p_{\min}\!=\!1~\mathrm{m}$ on the input force $F$ and on the output position $p$, respectively, while standard values are used for all other MPC settings in \eqref{eq:MPC} not mentioned here.

The design parameter $\theta$ has thus dimension $\ntheta=14$.

\subsection{Results}

\begin{figure}
\centering
\begin{subfigure}[b]{\textwidth}
   \includegraphics[width=1\linewidth]{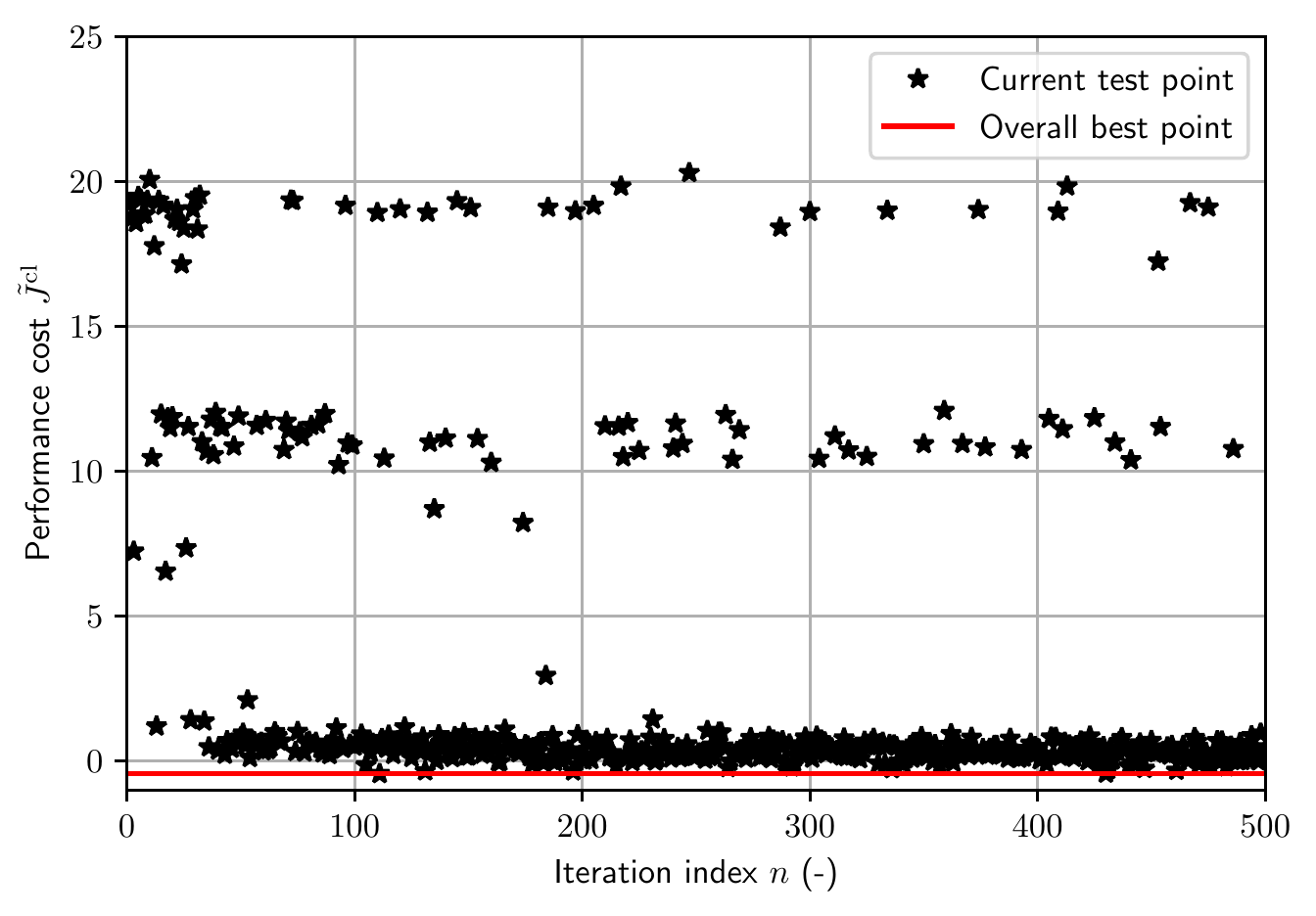}
\end{subfigure}

\begin{subfigure}[b]{\textwidth}
   \includegraphics[width=1\linewidth]{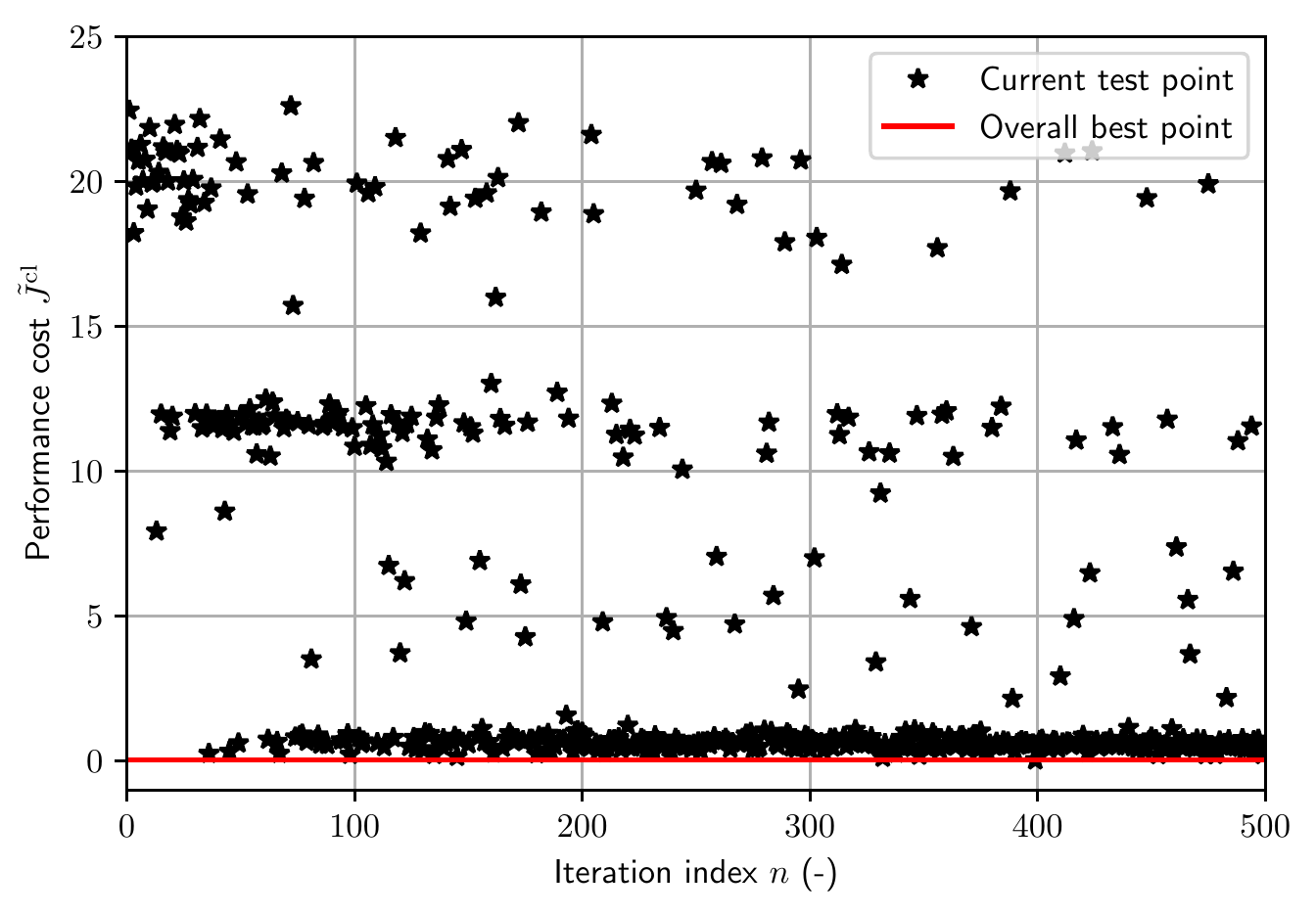}
\end{subfigure}
\caption{Performance cost $\JCt$ vs. iteration index $n$ of GLIS for experiments  run on the PC (top) and on the Raspberry PI (bottom).}
\label{fig:iterations}
\end{figure}
 

\begin{figure}
\centering
\begin{subfigure}[b]{\textwidth}
   \includegraphics[width=1\linewidth]{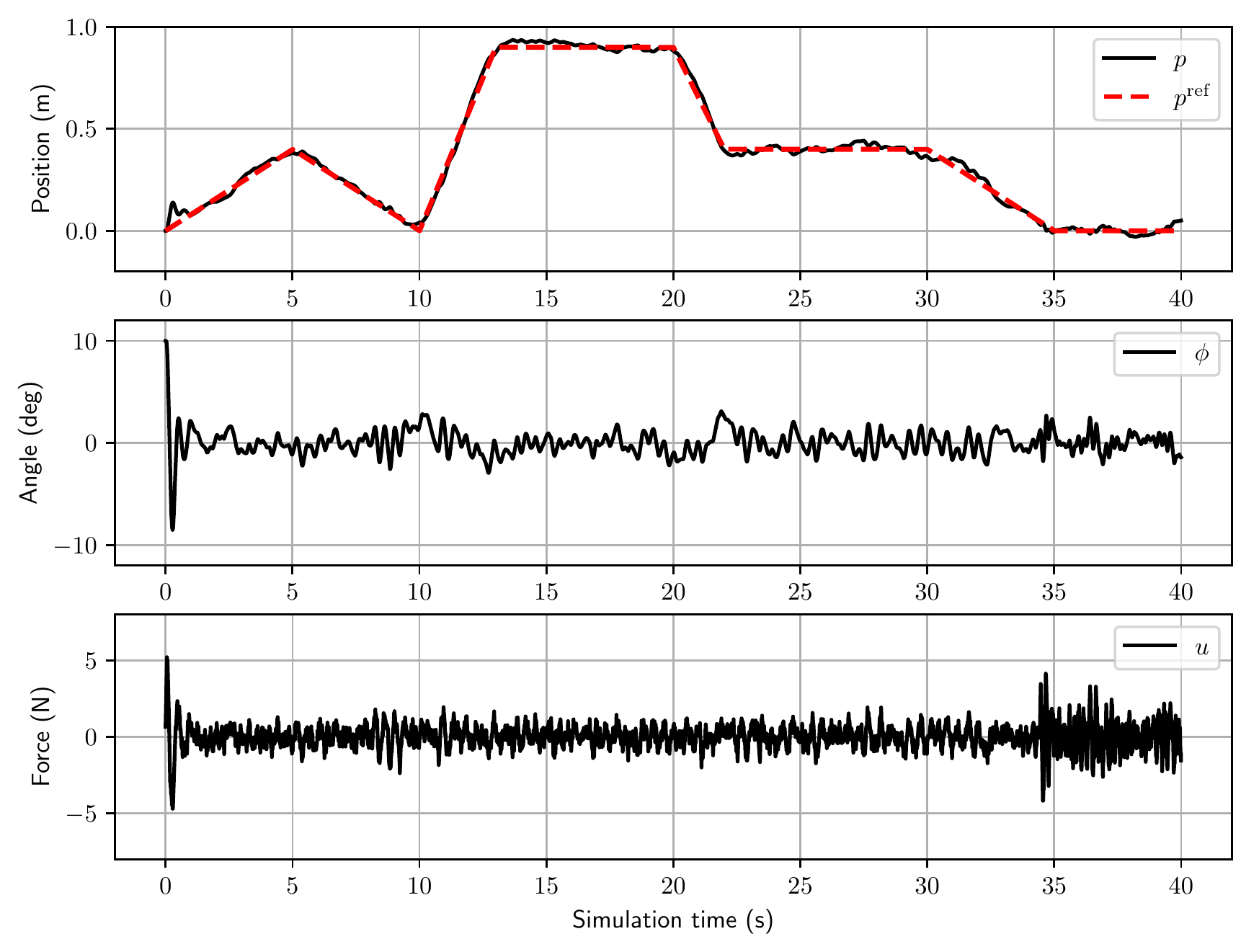}
\end{subfigure}

\begin{subfigure}[b]{\textwidth}
   \includegraphics[width=1\linewidth]{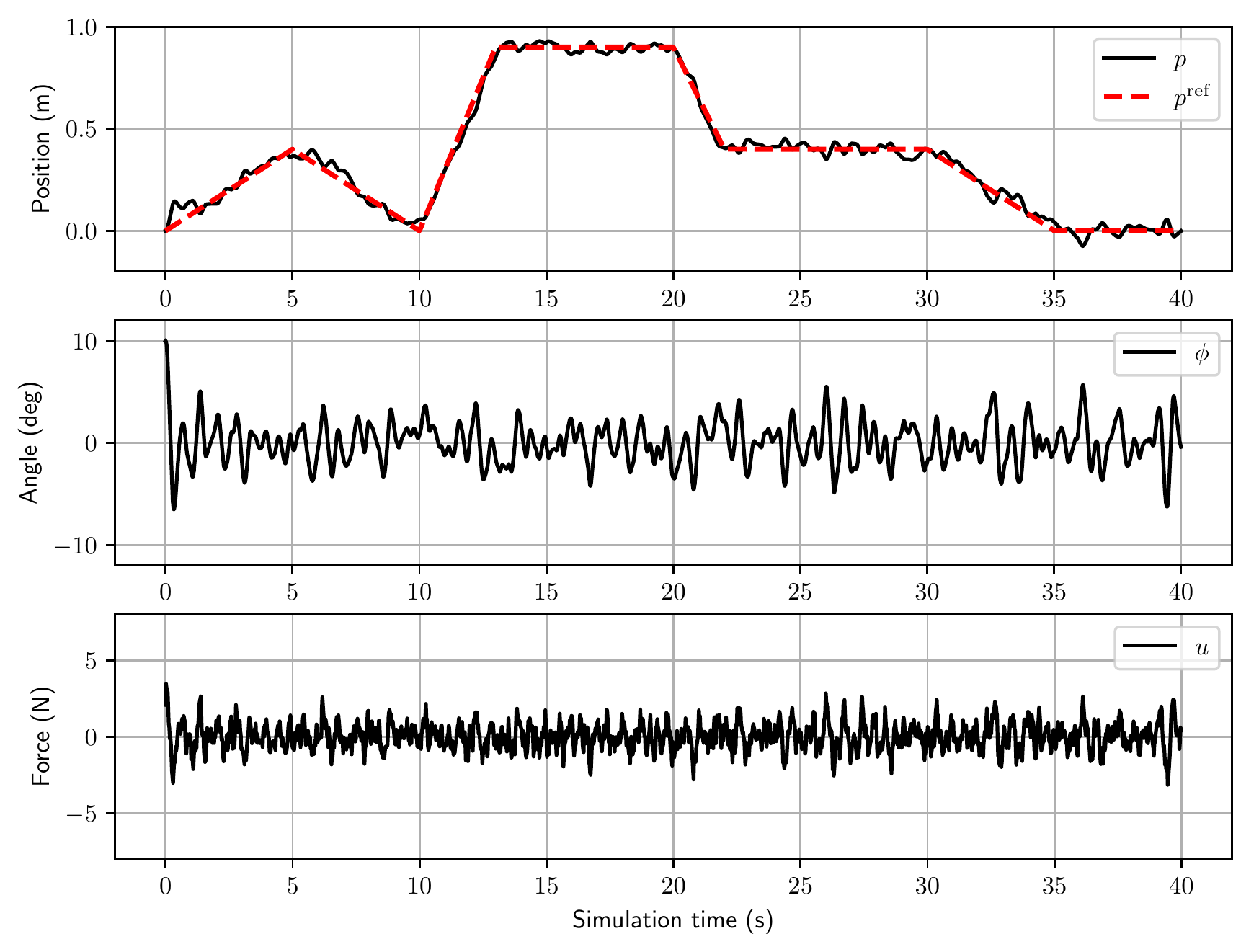}
\end{subfigure}
\caption{Performance achieved by the designed MPC  for experiments run on the PC (top panel) and on the Raspberry PI (bottom panel).}
\label{fig:timetraces}
\end{figure}

The global optimizer GLIS is run for $n_{\mathrm{max}} = 500$ iterations. The performance cost $\JCt$~\eqref{eq:JCL_example} \emph{vs.} the iteration index $n$ is shown in Fig.~\ref{fig:iterations} for the PC (top) and the Raspberry PI 3 (bottom). It can be observed that, after about $100$ iterations, the majority of the controller parameter configurations proposed by the optimization algorithm are concentrated in regions with low cost $\JCt$. 
This is more evident on the Raspberry PI 3 platform, where  the set of parameters satisfying the real-time constraint~\eqref{eq:TCALCcnst} is smaller. 
Persistent high values  $\JCt$ are due (correctly) to the exploration of the parameter space $\Theta$ promoted by the GLIS algorithm.

The obtained optimal performance index $\JCt$ after 500 iterations is slightly better on the PC (-0.44) than on the Raspberry PI (0.02), as expected. Indeed, certain MPC configurations characterized, \emph{e.g.}, by small sampling time and long prediction horizon may be feasible on the PC, but not on the Raspberry PI.

 Fig.~\ref{fig:timetraces} shows the time trajectories of position $p$, angle $\phi$, and force $F$ for the optimal MPC controller on the PC (top panel) and on the Raspberry PI (bottom panel), over a validation reference trajectory different from the one used for calibration. A slightly better performance for the MPC implementation on the PC can be appreciated both in terms of a tighter cart position tracking and a lower variance in the angle signal. 

  Analyzing the two  final MPC designs, we noticed that on the PC we have $\TMPC= 6~\mathrm{ms}$, while on the Raspberry PI we have $\TMPC= 22~\mathrm{ms}$.   The optimal solution found for the PC platform allows a faster loop update, and thus achieves superior trajectory tracking and noise rejection capabilities.
  On the other hand, a larger MPC time  is required on the Raspberry PI to guarantee real-time implementation. 

\section{Conclusions and follow-up}
\label{sec:conclusions}
We have presented an automated method to calibrate MPC parameters with a limited number of experiments.  Real-time implementation constraints are explicitly taken into account in the design in order to allow embedded implementation of the resulting controller. 

Future research will be devoted to find a parameterized  solution of the optimal MPC tuning knobs with respect to the reference trajectories, and to the analysis of the  generalization properties of the designed controllers against control objectives not considered in the calibration phase.


\bibliography{ms}             
                                                   
\end{document}